\documentclass[a4paper]{article}

\usepackage{INTERSPEECH2019}
\usepackage{array}
\usepackage{tabularx}
\usepackage{multirow}
\usepackage{cite}

\title{Spatial Pyramid Encoding with Convex Length Normalization for Text-Independent Speaker Verification}
\name{Youngmoon Jung$^1$, Younggwan Kim$^2$, Hyungjun Lim$^1$, Yeunju Choi$^1$, Hoirin Kim$^1$}
\address{
$^1$School of Electrical Engineering, KAIST, Daejeon, South Korea
\\$^2$Artificial Intelligence Lab, LG Electronics, Seoul, South Korea}
\email{dudans@kaist.ac.kr, younggwan.kim@lge.com, \{hyungjun.lim,wkadldppdy,hoirkim\}@kaist.ac.kr}

\begin{document}

\maketitle
\begin{abstract}
In this paper, we propose a new pooling method called spatial pyramid encoding (SPE) to generate speaker embeddings for text-independent speaker verification. We first partition the output feature maps from a deep residual network (ResNet) into increasingly fine sub-regions and extract speaker embeddings from each sub-region through a learnable dictionary encoding layer. These embeddings are concatenated to obtain the final speaker representation. 
The SPE layer not only generates a fixed-dimensional speaker embedding for a variable-length speech segment, but also aggregates the information of feature distribution from multi-level temporal bins. 
Furthermore, we apply deep length normalization by augmenting the loss function with ring loss. 
By applying ring loss, the network gradually learns to normalize the speaker embeddings using model weights themselves while preserving convexity, leading to more robust speaker embeddings. 
Experiments on the VoxCeleb1 dataset show that the proposed system using the SPE layer and ring loss-based deep length normalization outperforms both \textit{i}-vector and \textit{d}-vector baselines.

\end{abstract}
\noindent\textbf{Index Terms}: speaker verification, spatial pyramid encoding,  learnable dictionary encoding, ring loss, length normalization

\section{Introduction}

Speaker verification (SV) is the task of verifying a person's claimed identity based on his or her voice. 
Depending on the lexicon constraint on the spoken content, the SV systems can be classified into two categories, text-dependent speaker verification (TD-SV) and text-independent speaker verification (TI-SV). TD-SV requires the content of input speech to be fixed, while TI-SV operates on unconstrained speech.

The combination of \textit{i}-vector \cite{Dehak2011} and probabilistic linear discriminant analysis (PLDA) \cite{Ioffe2006} has been the dominant approach for TI-SV tasks \cite{Kenny2010, Garcia2011}. 
Recently, a deep neural network (DNN) trained for automatic speech recognition (ASR) was integrated into the \textit{i}-vector system, which improved the conventional Gaussian Mixture Model-Universal Background Model (GMM-UBM) based \textit{i}-vector system \cite{Kenny2014, Lei2014}. However, the use of the additional ASR-DNN drastically increases the computational complexity and also requires transcribed data for training.

Another deep learning-based approach is to extract speaker embeddings directly from a speaker discriminative network \cite{Variani2014, Chen2015, Zhang2018, NaLi2018, Huang2018}. 
In such systems, the network is trained to classify speakers in the training set, or to separate same-speaker and different-speaker utterance pairs.
After training, the utterance-level speaker embeddings (called \textit{d}-vectors) are obtained by aggregating the frame-level features extracted from the network.

Most \textit{d}-vector based SV systems use a pooling mechanism to map a variable-length segment to a fixed-dimensional embedding vector. 
Average pooling is the most common method to extract the utterance-level speaker representations \cite{Snyder2016, Li2017, Nagrani2017}.
Recently, some researchers have proposed more advanced pooling methods.
Snyder \textit{et al.} \cite{Snyder2017} introduced the statistics pooling layer in which the standard deviation is used as well as the mean. Okabe \textit{et al.} \cite{Okabe2018} combined the attention mechanism and the statistics pooling layer to propose attentive statistics pooling layer.
Zhang \textit{et al.} \cite{Zhang2018} proposed to replace the average pooling layer with the spatial pyramid pooling (SPP) layer \cite{He2014} to maintain spatial information by pooling in local spatial bins.
Cai \textit{et al.} \cite{Cai2018} applied the learnable dictionary encoding (LDE) scheme for extracting speaker embeddings. They imitated the process of encoding GMM supervectors within a deep learning framework. 
These approaches improved the performance over simple average pooling.

Once \textit{i}-vectors or \textit{d}-vectors are extracted, we usually apply length normalization for the speaker representations to have unit norm \cite{Garcia-Romero2011, Li2017}.
In \cite{Cai2018sec}, the authors introduced $L_2$-constraint based deep length normalization. They added an $L_2$-normalization layer followed by a scale layer to constrain the representations to lie on a hypersphere of a fixed radius. They showed that integrating this simple step in the training pipeline boosts the performance of speaker verification.

In this work, we propose a new pooling scheme, called spatial pyramid encoding (SPE). After the frame-level features are extracted from ResNet \cite{He2016}, we divide the feature maps of the last layer into uniform grids at different scales. Unlike using the average pooling operation in the SPP layer, we extract embeddings from each sub-region through the LDE layer.
The final speaker representation is produced by aggregating the embeddings from each sub-region.
Furthermore, we apply convex length normalization using ring loss \cite{Zheng2018} to normalize the speaker embedding. 
We show that ring loss-based deep length normalization performs better than the $L_2$-constraint based one.
 
In this paper, we first describe the \textit{d}-vector systems in Section 2. Section 3 reviews the related prior works. Section 4 presents our proposed methods. The experimental setup and results are described in Section 5 and Section 6, respectively. We conclude this work in Section 7.

\section{\textit{d}-vector systems}

We can classify \textit{d}-vector based SV systems according to the loss function used. The first one is based on the softmax loss defined in \cite{Liu2016} as the combination of a cross-entropy loss, a softmax function and the last fully connected layer \cite{Variani2014, Chen2015, Ravanelli2018}. In this system, a speaker classifier is trained to classify speakers in the training set. 
The softmax loss encourages the separability of speaker embeddings. However, the softmax loss is not sufficient to learn the discriminative embedding with a large margin, and more researchers began to explore discriminative loss functions for enhanced generalization ability. 

Another type of system is based on the triplet loss \cite{Zhang2018} which enhances the intra-class compactness and inter-class separability, leading to better generalization ability. It minimizes the distance between embedding pairs from the same speaker and maximizes the distance between pairs from different speakers.
A drawback is that it requires the careful selection of triplets of samples, which is time-consuming and performance-sensitive.

To circumvent the triplet-wise computation and learn more discriminative representations, the center loss \cite{Wen2016} and angular softmax (A-softmax) loss \cite{Liu2017} are applied to SV tasks, respectively \cite{NaLi2018, Huang2018}. 
The center loss 
minimizes the Euclidean distance between the embeddings and the corresponding class centroids. 
The angular softmax loss introduces an angular margin into the softmax loss through the designing of a sophisticated differentiable angular distance function. The hyperparameter $m$ controls the size of the angular margin. Large $m$ gives more stringent constraint on the distribution of the deep embeddings and enforces a larger angular margin between classes.

For all the systems mentioned above, the frame-level features are extracted from the speaker discriminative network.
Then, the \textit{d}-vector is obtained by a pooling layer that aggregates the frame-level features across time. 
The speaker-dependent \textit{d}-vector for each enrollment speaker is stored after the \textit{d}-vector is divided by its $L_2$-norm for length normalization.
Finally, scoring between enrollment and test \textit{d}-vector is performed using either the cosine distance or PLDA.

\section{Prior works}


\subsection{Learnable dictionary encoding layer}

Cai \textit{et al.} \cite{Cai2018} employed the learnable dictionary encoding (LDE) layer \cite{Zhang2017} for speaker recognition. 
The LDE layer acts as a pooling layer integrated on top of convolutional layers, which ports the entire dictionary learning and encoding pipeline into a single model. It accepts variable-length inputs and produces fixed-length speaker embeddings. 
We assume that frame-level features are distributed in $C$ codewords and the LDE layer learns a dictionary, a set of codewords.
This is essentially the same as the conventional GMM supervector.


The LDE layer considers an input feature map with the shape of $H \times W \times D$ as a set of $D$-dimensional input features $X = \{x_1,	...,x_L\}$, where $L$ is the total number of features given by $H \times W$,
which learns an inherent codebook $\mu = \{\mu_1,...,\mu_C\}$ containing $C$ number of codewords and a set of smoothing factor of the codewords $S = \{s_1,...,s_C\}$. The residual encoding $e_c$ for codeword $\mu_c$ is generated by aggregating the residuals with soft-assignment weights:
\begin{equation}
e_{c} = \sum_{t=1}^L e_{tc} = \frac{\sum_{t=1}^L w_{tc}r_{tc}}{L}\,,
\end{equation}
where the residuals are given by $r_{tc} = x_t - \mu_c$. The assigning weight is given by a softmax function as follows:
\begin{equation}
w_{tc} = \frac{exp(-s_{c}\|r_{tc}\|^2)}{\sum_{m=1}^C exp(-s_m\|r_{tm}\|^2)}\,.
\end{equation}

The LDE layer concatenates the residual encoding vectors, generating a fixed-length representation $E = \{e_1,...,e_C\}$ (independent of the number of input features $L$). 
The resulting vector $E$ has the same role as the supervector in the GMM supervector approach. 
Finally, this supervector is projected to a lower dimension to obtain the final embedding through an additional fully connected (FC) layer. This projection has the same role as the total variability matrix of the \textit{i}-vector system.


\subsection{$L_2$-constraint based deep length normalization}

\begin{figure}[!htb]
  \vspace{-2.8cm}
  \centerline{\includegraphics[width=13cm]{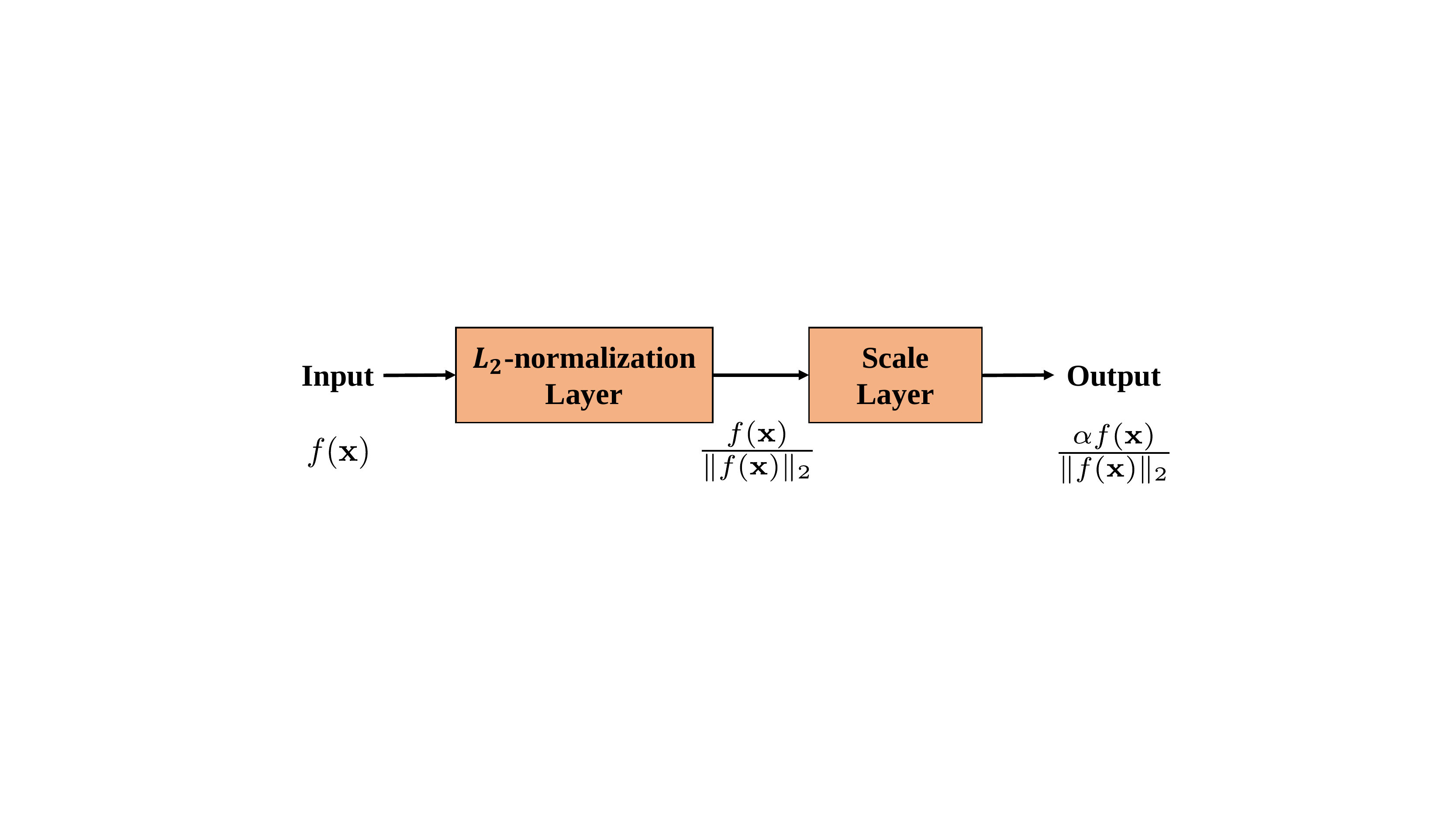}}
  \vspace{-3cm}
  \caption{$L_2$-constraint based deep length normalization}
  \vspace{-0.05cm}
  \label{L2_constraint}
  \vspace{-0.05cm}
\end{figure}

Cai \textit{et al.} \cite{Cai2018sec} applied an $L_2$-constraint \cite{Ranjan2017} to the speaker embedding during training. As shown in Figure \ref{L2_constraint}, they added an $L_2$-normalization layer followed by a scale layer to constrain the speaker embedding to lie on a hypersphere of a fixed radius. 

This module is added just after the penultimate layer of the network which is the pooling layer. 
The $L_2$-normalization layer normalizes the input speaker embedding $f(\mathbf x)$ to a unit vector. The scale layer scales the unit-length embedding vector into a fixed radius given by the parameter $\alpha$. They showed that this simple step in the training pipeline boosts the performance of speaker verification systems.

\begin{figure*}[t]
  \centerline{\includegraphics[width=17cm]{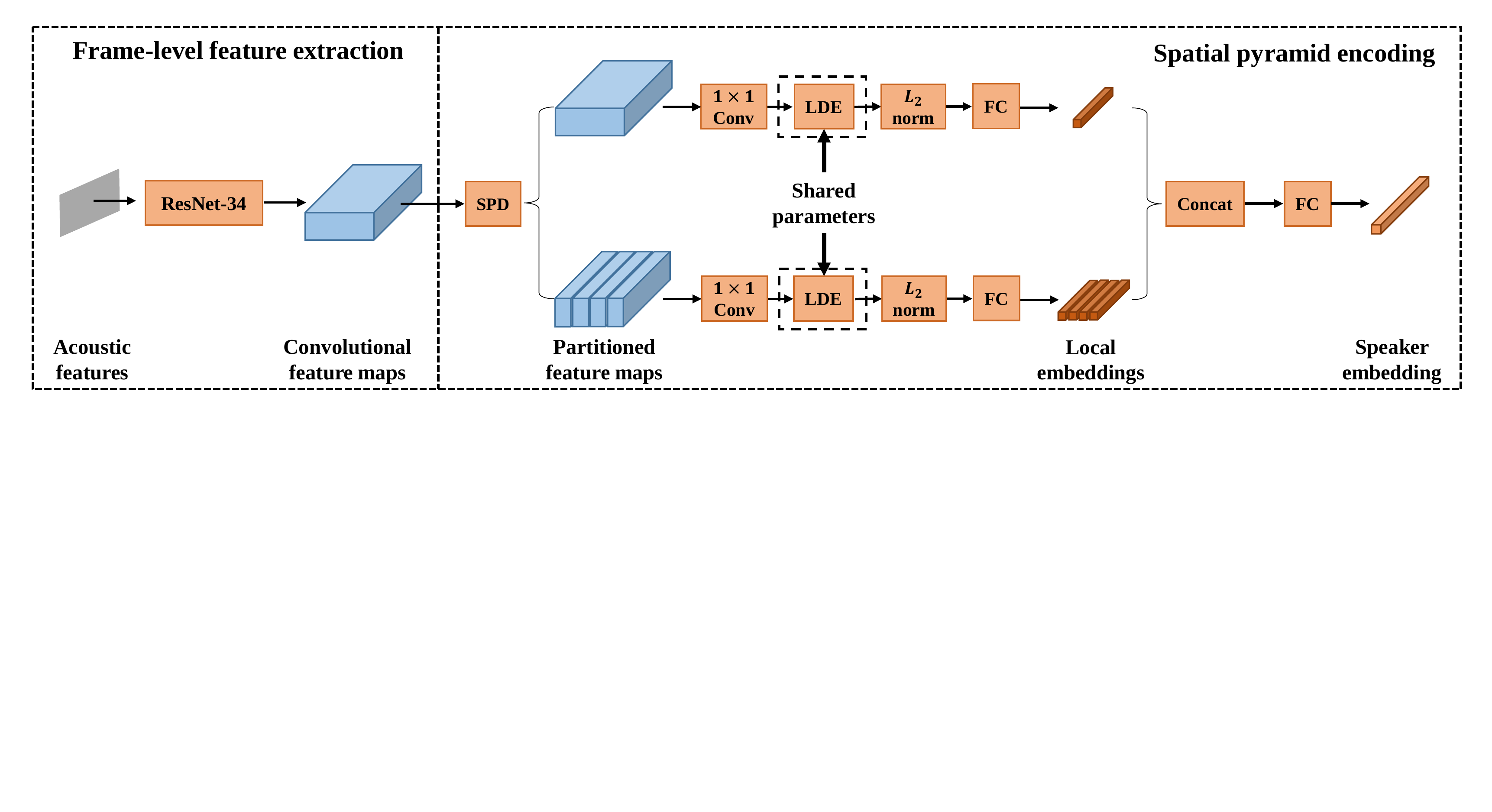}}
  \vspace{-4.75cm}
  \caption{Overview of the proposed spatial pyramid encoding (SPE) layer.}
  \vspace{-0.4cm}
  \label{SPE}
\end{figure*}

\section{Proposed approaches}
\subsection{Spatial pyramid encoding layer}

\begin{table}[t]
\centering
\begin{footnotesize}
\renewcommand{\arraystretch}{1.2}
\caption{The architecture of the frame-level feature extractor based on 34-layer ResNet \cite{He2016}. The input size is $64 \times T$.}
\label{architecture}
\begin{tabular}{c|c|cll}
\cline{1-3}
stage & output size    & ResNet-34 \\ \cline{1-3}
conv1      & $64 \times T \times 32$    & $7 \times 7, 32$, stride $1$                                              \\ \cline{1-3}
conv2       & $64 \times T \times 32$    & $\left[ \begin{array}{cc} 3 \times 3, 32  \\ 3 \times 3, 32 \end{array}\right]$ $\times$ $3$ \\ \cline{1-3}
conv3       & $32 \times T/2 \times 64$  & $\left[ \begin{array}{cc} 3 \times 3, 64  \\ 3 \times 3, 64 \end{array}\right]$ $\times$ $4$ \\ \cline{1-3}
conv4       & $16 \times T/4 \times 128$ & $\left[ \begin{array}{cc} 3 \times 3, 128  \\ 3 \times 3, 128 \end{array}\right]$ $\times$ $6$ \\ \cline{1-3}
conv5       & $8 \times T/8 \times 256$  & $\left[ \begin{array}{cc} 3 \times 3, 256  \\ 3 \times 3, 256 \end{array}\right]$ $\times$ $3$ \\ \cline{1-3}

\end{tabular}
\end{footnotesize}
\vspace{-0.25cm}
\end{table}


Figure \ref{SPE} shows the proposed pooling layer, called the spatial pyramid encoding (SPE) layer.
First, the 34-layer ResNet is used to extract frame-level features from utterances, which has been widely used in previous studies \cite{Li2017, Cai2018, Cai2018sec, Le2018}. 
The architecture is described in Table \ref{architecture}.
The ResNet takes log Mel-filterbank (Fbank) features of size $64 \times T \times 1$ and outputs frame-level features of size $8 \times T/8 \times 256$.
The resulting feature maps are fed into the SPE layer and then aggregated into a single, utterance-level speaker representation. 

The SPE method includes three steps.
In the first step, the input feature maps are divided into increasingly finer sub-regions along the time axis, forming a pyramid of sub-feature maps. This operation is called the spatial pyramid division (SPD). In this work, we apply the pyramids with two levels \{$1 \times 1$, $1 \times 4$\} (totally 5 bins).
Subsequently, a $1 \times 1$ convolutional layer is used for each bin, reducing the number of channels from 256 to 64. After that, we extract speaker embeddings from each bin through the LDE layer with 64 codewords, followed by $L_2$-normalization and an FC layer. This FC layer reduces the dimension of the embeddings from 4,096 (= 64 $\times$ 64) to 256.
Here, the LDE layer is shared across all bins.
At last, all the local embeddings are concatenated and passed through an FC layer with 256 neurons to form the final speaker embedding. 


The SPE layer can be viewed as a combination of the LDE layer and spatial pyramid pooling (SPP) \cite{He2014} layer. SPP (also known as spatial pyramid matching or SPM \cite{Lazebnik2006}), as an extension of the bag-of-words (BoW) model \cite{Sivic2003}, has been widely used in the computer vision community.
It partitions an image into several segments in different scales, then computes the BoW histograms \cite{Lazebnik2006} or GMM supervectors \cite{Kamishima2013} of local features in each segment. The resulting vectors for all the segments are concatenated to form a high dimensional vector representation of the image. SPP enables us to incorporate the spatial information of feature vectors. He \textit{et al.} \cite{He2014} proposed SPP-net in which the SPP layer is used to replace the last pooling layer of the convolutional neural network (CNN). Later, the SPP layer was applied to speaker verification tasks \cite{Zhang2018}. 
In the SPP layer, the last convolutional feature maps are divided into sub-regions, and then average pooling is applied to each sub-region.

The proposed SPE layer replaces the simple average pooling operation of the SPP layer with the LDE operation which is found to perform better for speaker verification task in \cite{Cai2018}.
Therefore, the SPE layer can be seen as the extension of the SPP layer. 
At the same time, we can also view the SPE layer as the extension of the LDE layer.
The descriptive power of the LDE layer is limited because it discards the temporal information of local CNN features. 
This motivates us to combine temporal information with the LDE layer.
The SPE layer enhances the LDE layer by taking the temporal information into consideration at both local and global scales.

\subsection{Ring loss-based deep length normalization}

The $L_2$-constraint based deep length normalization explained in Section 3.2 uses the norm constraint right before the softmax loss. However, according to \cite{Zheng2018}, such a direct approach through the hard normalization operation results in a non-convex formulation. It results in local minima generated by the loss function itself and leads to difficulties in optimization. It is important to preserve convexity in loss functions for more effective minimization of the loss given that the network optimization itself is non-convex. To deal with this issue, we apply ring loss \cite{Zheng2018} that normalizes deep speaker embeddings through a convex augmentation of the primary loss function (such as softmax loss \cite{Liu2016} or A-softmax loss \cite{Liu2017}).
To the best of our knowledge, this is the first work to apply ring loss to speaker verification systems. 
Ring loss $L_R$ is defined as
\begin{equation}
L_R = \frac{1}{m} \sum_{i=1}^m \Big(\frac{\|f(\mathbf x_i)\|_2-R}{\mathop{\mathbb{E}} \left[\|f(\mathbf x)\|_2\right]}\Big)^2\,,
\end{equation}
where $f(\mathbf x_i)$ is the speaker embedding for the sample $\mathbf x_i$. Here, $R$ is the target norm value which is learned during training, $m$ is the batch size, and $\mathop{\mathbb{E}} \left[\|f(\mathbf x)\|_2\right] = \frac{1}{m}\sum_{i=1}^m(\|f(\mathbf x_i)\|_2)$, which is the average $L_2$-norm of the input embedding vectors for each mini-batch. The loss encourages the norm of the embeddings being value $R$ (a learned parameter) rather than explicit enforcing through a hard normalization operation as in the $L_2$-constraint based method. 
The total objective function is formulated as
\begin{equation}
L = L_P + \lambda L_R\,,
\end{equation}
where $L_P$ is the primary loss function. A scalar $\lambda$ is used for balancing the two loss functions, which is the only hyperparameter in ring loss. 
In this work, the $\mathop{\mathbb{E}} \left[\|f(\mathbf x)\|_2\right]$ obtained from the first iteration of training is used as the initial value of $R$. 

\section{Experimental setup}

\subsection{Datasets}
In this paper, we train our models on the VoxCeleb1 dataset \cite{Nagrani2017}.
The VoxCeleb1 dataset is a large scale text-independent speaker recognition dataset, which contains over 140,000 utterances from 1,251 distinct celebrities, in real-world conditions.
For the speaker verification task, there are a total of 1,211 speakers in the development set and the rest 40 speakers are reserved as the test set. For further details, please refer to \cite{Nagrani2017}.

We report the equal error rate (EER) and the minimum detection cost function (DCF) \cite{Hansen2015} at $P_{target}$ = 0.01 and $P_{target}$ = 0.001. Verification trials are scored using cosine distance.

\subsection{Implementation details}

The input acoustic features are 64-dimensional Fbank features with a frame-length of 25 ms, which are mean-normalized over a sliding window of up to 3 s. 
Both voice activity detection (VAD) and data augmentation are not applied in the systems.

For each training step, an integer $T$ is randomly selected within [300, 500] interval, and the input utterance is cropped or extended to $T$ frames. Thus, the input size of the ResNet-34 model is $64 \times T$ as shown in Table \ref{architecture}.
After training, the entire utterance is evaluated at once in the testing stage. The 256-dimensional speaker embeddings are extracted from a pooling layer. When deep length normalization is applied in training, we do not need an additional length normalization step in testing.

The models are implemented with PyTorch \cite{Paszke2017} and optimized by stochastic gradient descent with momentum 0.9. The mini-batch size is 64, and the weight decay parameter is 0.0001. We use the same learning rate schedule as in \cite{Cai2018} with the initial learning rate of 0.1.

In LDE layers, the number of codewords $C$ is 64. 
We use the angular margin $m$ = 4 for A-softmax loss. 
The hyperparameter for ring loss $\lambda$ is set to 1.

\section{Results}

\begin{table}[t]
\centering
\caption{The performance comparison of different pooling methods. The  softmax loss with ring loss is used. ``2D'' denotes that the spatial pyramid division (SPD) of \{$1 \times 1$, $2 \times 2$\} is applied as in \cite{Zhang2018}, and ``1D'' denotes that the SPD of \{$1 \times 1$, $1 \times 4$\} is applied as explained in Section 4.1.}
\label{pooling}
\begin{footnotesize}
\begin{tabular}{cccc}
\hline
Pooling         & EER (\%)          & DCF $10^{-2}$     & DCF $10^{-3}$    \\ \hline
TAP             & 4.62          & 0.460         & 0.581         \\
LDE             & 4.33          & 0.435         & 0.549         \\
2D-SPP          & 4.59          & 0.452         & 0.573         \\
1D-SPP          & 4.50          & 0.447         & 0.564         \\
2D-SPE          & 4.29          & 0.428         & 0.534         \\
\textbf{1D-SPE}   & \textbf{4.20} & \textbf{0.422} & \textbf{0.528} \\ \hline
\end{tabular}
\end{footnotesize}
\end{table}

\begin{table}[t]
\centering
\renewcommand{\arraystretch}{1.3}
\caption{The performance comparison of different deep length normalization methods. Temporal average pooling is used. SM denotes the softmax loss, ASM denotes the A-softmax loss, $L_2$-Cons denotes $L_2$-constraint based deep length normalization, and finally ``+ Ring'' denotes ring loss augmentation.} 
\label{loss_and_norm}
\begin{scriptsize}
\begin{tabular}{cccccc}
\hline
Loss $\&$ Norm         & $R$                 & EER (\%)           & DCF $10^{-2}$ & DCF $10^{-3}$   \\ \hline
SM                     & -                 & 6.87          & 0.538         & 0.708 \\
$L_2$-Cons SM          & 12 (F)            & 4.83          & 0.479         & 0.572           \\
$L_2$-Cons SM          & 24.1 (L)          & 5.13          & 0.498         & 0.601           \\
SM + Ring              & 20.5 (L)          & 4.62          & 0.460         & 0.581           \\
ASM                    & -                 & 4.88          & 0.499         & 0.597           \\
$L_2$-Cons ASM         & 30 (F)            & 4.69          & 0.478         & 0.584           \\
$L_2$-Cons ASM         & 28.3 (L)          & 4.73          & 0.475         & 0.594           \\ 
\textbf{ASM + Ring}    & \textbf{24.8 (L)} & \textbf{4.41} & \textbf{0.451} & \textbf{0.559} \\ \hline
\vspace{-0.45cm}
\end{tabular}
\end{scriptsize}
\end{table}

\begin{table}[t]
\centering
\renewcommand{\arraystretch}{1.3}
\caption{Comparison of the proposed and state-of-the-art systems. SAP denotes self-attentive pooling and SP denotes statistics pooling. Other abbreviations are the same as in Table \ref{loss_and_norm}.}
\label{comparison}
\begin{scriptsize}
\begin{tabular}{ccccc}
\hline
Systems                               & Loss $\&$ Norm   & Pooling       & Scoring   & EER (\%)      \\ \hline
\textit{i}-vector \cite{Shon2018}     & -                & -             & PLDA      & 5.4        \\
VGG-M \cite{Nagrani2017}                & Contrastive      & TAP           & Cosine      & 7.8           \\
VGG (1D) \cite{Shon2018}          & SM              & SP       & PLDA      & 5.3        \\
VGG-13 \cite{Yadav2018}      & Center           & TAP          & Cosine      & 4.9        \\
ResNet-34 \cite{Cai2018}     & ASM              & TAP          & PLDA      & 4.46          \\ 
ResNet-34 \cite{Cai2018}     & ASM              & SAP          & PLDA      & 4.40          \\ 
ResNet-34 \cite{Cai2018}     & ASM              & LDE          & PLDA      & 4.48          \\ 
ResNet-34 \cite{Cai2018sec}  & $L_2$-Cons SM    & TAP          & PLDA      & 4.74          \\ 
\textbf{Proposed}                     & \textbf{ASM + R} & \textbf{SPE} & \textbf{Cosine}    & \textbf{4.03} \\ \hline
\vspace{-0.45cm}
\end{tabular}
\end{scriptsize}
\end{table}


\subsection{Comparison of pooling methods}
Table \ref{pooling} compares the performance of different pooling methods.
We use the softmax loss with ring loss-based deep length normalization for all cases.
As in \cite{Cai2018}, temporal average pooling (TAP) is essentially the same as global average pooling, which takes the average over all elements in the 2D feature map.
1D-SPE is our proposed SPE layer, in which the SPD is applied along the time axis as explained in Section 4.1.

Both the SPP and LDE layers yield better performance than the simple TAP layer. 
They provide relative improvements of 2.6\% and 6.3\% in EER over the TAP layer, respectively. 
In both the SPP and SPE layers, the 1D-SPD performs better than the 2D-SPD.
The best result (EER = 4.20\%, DCF $10^{-2}$ = 0.422, DCF $10^{-3}$ = 0.528) is obtained when the 1D-SPE layer is used. 
We can see that our proposed SPE layer (1D-SPE) performs better than both the SPP and LDE layers, achieving relative improvements of 6.7\% and 3.0\% in EER, respectively. 

\subsection{Comparison of deep length normalization methods}
In Table \ref{loss_and_norm}, we compare the performance of different deep length normalization methods. 
In the second column, we present the target norm value $R$ that we would like the speaker embeddings to be normalized to. 
In the $L_2$-constraint based method ($L_2$-Cons), $R$ is equal to $\alpha$ defined in Section 3.2. ``(F)'' denotes that a fixed optimal $R$ value is used, and ``(L)'' denotes that the parameter $R$ is learned by the network rather than fixed.

The softmax loss is used in the first four entries, and the A-softmax loss is used in the last four entries.
We observe that applying deep length normalization leads to performance improvement. For example, using the softmax loss with the ring loss (SM + Ring) shows a relative improvement of 32.8\% in EER over using the softmax loss without the ring loss (SM).
Furthermore, we can see that the proposed ring loss-based deep length normalization performs better than the $L_2$-constraint based approach.
When using the A-softmax loss, the ring loss achieves a relative improvement of 6.0\% in EER over the $L_2$-Cons with $R$ = 30.
The best result (EER = 4.41\%, DCF $10^{-2}$ = 0.451, DCF $10^{-3}$ = 0.559) is obtained when the A-softmax loss function is used with ring loss-based deep length normalization.

\subsection{Comparison with recent methods}
In Table \ref{comparison}, we compare our proposed system with recently reported SV systems in terms of EER. 
For fair comparisons, we do not include systems that are trained on a larger dataset such as VoxCeleb2 \cite{Chung2018}, or that use data augmentation such as \cite{Okabe2018}. 
The \textit{i}-vector + PLDA system \cite{Shon2018} uses 2,048 Gaussian components.
VGG-M \cite{Nagrani2017} is trained using contrastive loss with the TAP layer.
VGG (1D) \cite{Shon2018} uses a 1D-CNN instead of a 2D-CNN, and the statistics pooling layer. VGG-13 \cite{Yadav2018} is trained under the joint supervision of softmax loss and center loss. 
The ResNet-34 based systems in \cite{Cai2018} use the TAP, SAP, and LDE layer, respectively. The ResNet-34 based system in \cite{Cai2018sec} applies $L_2$-constraint based deep length normalization.

The proposed system uses the SPE layer and A-softmax loss with ring loss. 
We obtain an EER of 4.03\%, a DCF $10^{-2}$ of 0.402, and a DCF $10^{-3}$ of 0.492. 
Our model outperforms all other state-of-the-art systems, including \textit{i}-vector and other \textit{d}-vector systems. It yields relative improvements of 25.4\% and 8.4\% over the \textit{i}-vector system and ResNet-34 + SAP (which shows the best performance among the baselines), respectively. 

\section{Conclusions}

In this paper, we proposed spatial pyramid encoding to extract \textit{d}-vectors for TI-SV. This method achieved better results than the LDE and SPP method.
Furthermore, we applied ring loss-based deep length normalization, and it performed better than the existing $L_2$-constraint based one. 
On the VoxCeleb1 dataset, our system using the SPE layer and ring loss obtained better performance than the state-of-the-art \textit{i}-vector and \textit{d}-vector baselines.
In the future, we will explore how to automatically divide the feature maps of CNNs in the SPE layer. 

\section{Acknowledgements}

This material is based upon work supported by the Ministry of Trade, Industry and Energy (MOTIE, Korea) under Industrial Technology Innovation Program (No.10063424, Development of distant speech recognition and multi-task dialog processing technologies for in-door conversational robots).

\bibliographystyle{IEEEtran}

\bibliography{mybib}

\begin{thebibliography}{10}
\providecommand{\url}[1]{#1}
\csname url@samestyle\endcsname
\providecommand{\newblock}{\relax}
\providecommand{\bibinfo}[2]{#2}
\providecommand{\BIBentrySTDinterwordspacing}{\spaceskip=0pt\relax}
\providecommand{\BIBentryALTinterwordstretchfactor}{4}
\providecommand{\BIBentryALTinterwordspacing}{\spaceskip=\fontdimen2\font plus
\BIBentryALTinterwordstretchfactor\fontdimen3\font minus
  \fontdimen4\font\relax}
\providecommand{\BIBforeignlanguage}[2]{{%
\expandafter\ifx\csname l@#1\endcsname\relax
\typeout{** WARNING: IEEEtran.bst: No hyphenation pattern has been}%
\typeout{** loaded for the language `#1'. Using the pattern for}%
\typeout{** the default language instead.}%
\else
\language=\csname l@#1\endcsname
\fi
#2}}
\providecommand{\BIBdecl}{\relax}
\BIBdecl

\bibitem{Dehak2011}
N.~Dehak, P.~J. Kenny, R.~Dehak, P.~Dumouchel, and P.~Ouellet, ``Front-end
  factor analysis for speaker verification,'' \emph{IEEE Transactions on Audio,
  Speech and Language Processing}, vol.~19, no.~4, pp. 788--798, 2011.

\bibitem{Ioffe2006}
S.~Ioffe, ``Probabilistic linear discriminant analysis,'' in \emph{Proceedings
  of European Conference on Computer Vision (ECCV)}, 2006, pp. 531--542.

\bibitem{Kenny2010}
P.~Kenny, ``Bayesian speaker verification with heavy tailed priors,'' in
  \emph{Proceedings of Odyssey Speaker and Language Recognition Workshop},
  2010, p.~14.

\bibitem{Garcia2011}
D.~Garcia-Romero and C.~Espy-Wilson, ``Analysis of ivector length normalization
  in speaker recognition systems,'' in \emph{Proceedings of Interspeech}, 2011,
  pp. 249--252.

\bibitem{Kenny2014}
P.~Kenny, V.~Gupta, T.~Stafylakis, P.~Ouellet, and J.~Alam, ``Deep neural
  networks for extracting baum-welch statistics for speaker recognition,'' in
  \emph{Proceedings of Odyssey Speaker and Language Recognition Workshop},
  2014, pp. 293--298.

\bibitem{Lei2014}
Y.~Lei, N.~Scheffer, L.~Ferrer, and M.~McLaren, ``A novel scheme for speaker
  recognition using a phonetically-aware deep neural network,'' in
  \emph{Proceedings of the IEEE International Conference on Acoustics, Speech
  and Signal Processing (ICASSP)}, 2014, pp. 1695--1699.

\bibitem{Variani2014}
E.~Variani, X.~Lei, E.~McDermott, I.~Moreno, and J.~GonzalezDominguez, ``Deep
  neural networks for small footprint text-dependent speaker verification,'' in
  \emph{Proceedings of the IEEE International Conference on Acoustics, Speech
  and Signal Processing (ICASSP)}, 2014, pp. 4052--4056.

\bibitem{Chen2015}
Y.~Chen, I.~Lopez-Moreno, T.~N. Sainath, M.~Visontai, R.~Alvarez, and
  C.~Parada, ``Locally-connected and convolutional neural networks for small
  footprint speaker recognition,'' in \emph{Proceedings of Interspeech}, 2015,
  pp. 1136--1140.

\bibitem{Zhang2018}
C.~Zhang, K.~Koishida, and J.~Hansen, ``Text-independent speaker verification
  based on triplet convolutional neural network embeddings,'' \emph{IEEE/ACM
  Transactions on Audio Speech and Language Processing}, vol.~26, no.~9, pp.
  1633--1644, 2018.

\bibitem{NaLi2018}
N.~Li, D.~Tuo, D.~Su, Z.~Li, and D.~Yu, ``Deep discriminative embeddings for
  duration robust speaker verification,'' in \emph{Proceedings of Interspeech},
  2018, pp. 2262--2266.

\bibitem{Huang2018}
Z.~Huang, S.~Wang, and K.~Yu, ``Angular softmax for short duration
  text-independent speaker verification,'' in \emph{Proceedings of
  Interspeech}, 2018, pp. 3623--3627.

\bibitem{Snyder2016}
D.~Snyder, P.~Ghahremani, D.~Povey, D.~Garcia-Romero, Y.~Carmiel, and
  S.~Khudanpur, ``Deep neural network based speaker embeddings for end-to-end
  speaker verification,'' in \emph{Proceedings of Spoken Language Technology
  Workshop (SLT)}, 2016, pp. 165--170.

\bibitem{Li2017}
C.~Li, X.~Ma, B.~Jiang, X.~Li, X.~Zhang, X.~Liu, Y.~Cao, A.~Kannan, and Z.~Zhu,
  ``Deep speaker: An end-to-end neural speaker embedding system,'' \emph{arXiv
  preprint arXiv:1705.02304}, 2017.

\bibitem{Nagrani2017}
A.~Nagrani, J.~S. Chung, and A.~Zisserman, ``Voxceleb: A large-scale speaker
  identification dataset,'' in \emph{Proceedings of Interspeech}, 2017, pp.
  2616--2620.

\bibitem{Snyder2017}
D.~Snyder, D.~Garcia-Romero, D.~Povey, and S.~Khudanpur, ``Deep neural network
  embeddings for text-independent speaker verification,'' in \emph{Proceedings
  of Interspeech}, 2017, pp. 999--1003.

\bibitem{Okabe2018}
K.~Okabe, T.~Koshinaka, and K.~Shinoda, ``Attentive statistics pooling for deep
  speaker embedding,'' in \emph{Proceedings of Interspeech}, 2018, pp.
  2252--2256.

\bibitem{He2014}
K.~He, X.~Zhang, S.~Ren, and J.~Sun, ``Spatial pyramid pooling in deep
  convolutional networks for visual recognition,'' in \emph{Proceedings of
  European Conference on Computer Vision (ECCV)}, 2014, pp. 346--361.

\bibitem{Cai2018}
W.~Cai, J.~Chen, and M.~Li, ``Exploring the encoding layer and loss function in
  end-to-end speaker and language recognition system,'' in \emph{Proceedings of
  Odyssey Speaker and Language Recognition Workshop}, 2018, pp. 74--81.

\bibitem{Garcia-Romero2011}
D.~Garcia-Romero and C.~Y. Espy-Wilson, ``Analysis of i-vector length
  normalization in speaker recognition systems,'' in \emph{Proceedings of
  Interspeech}, 2011, pp. 249--252.

\bibitem{Cai2018sec}
W.~Cai, J.~Chen, and M.~Li, ``Analysis of length normalization in end-to-end
  speaker verification system,'' in \emph{Proceedings of Interspeech}, 2018,
  pp. 3618--3622.

\bibitem{He2016}
K.~He, X.~Zhang, S.~Ren, and J.~Sun, ``Deep residual learning for image
  recognition,'' in \emph{Proceedings of Computer Vision and Pattern
  Recognition (CVPR)}, 2016, pp. 770--778.

\bibitem{Zheng2018}
Y.~Zheng, D.~K. Pal, and M.~Savvides, ``Ring loss: Convex feature normalization
  for face recognition,'' in \emph{Proceedings of Computer Vision and Pattern
  Recognition (CVPR)}, 2018, pp. 5089--5097.

\bibitem{Liu2016}
W.~Liu, Y.~Wen, Z.~Yu, and M.~Yang, ``Large-margin softmax loss for
  convolutional neural networks,'' in \emph{Proceedings of International
  Conference on Machine Learning (ICML)}, 2016, pp. 507--516.

\bibitem{Ravanelli2018}
M.~Ravanelli and Y.~Bengio, ``Speaker recognition from raw waveform with
  sincnet,'' in \emph{Proceedings of Spoken Language Technology Workshop
  (SLT)}, 2018.

\bibitem{Wen2016}
Y.~Wen, K.~Zhang, Z.~Li, and Y.~Qiao, ``A discriminative feature learning
  approach for deep face recognition,'' in \emph{Proceedings of European
  Conference on Computer Vision (ECCV)}, 2016, pp. 499--515.

\bibitem{Liu2017}
W.~Liu, Y.~Wen, Z.~Yu, M.~Li, B.~Raj, and L.~Song, ``Sphereface: Deep
  hypersphere embedding for face recognition,'' in \emph{Proceedings of
  Computer Vision and Pattern Recognition (CVPR)}, 2017, pp. 212--220.

\bibitem{Zhang2017}
H.~Zhang, J.~Xue, and K.~Dana, ``Deep ten: Texture encoding network,'' in
  \emph{Proceedings of Computer Vision and Pattern Recognition (CVPR)}, 2017,
  pp. 708--717.

\bibitem{Ranjan2017}
R.~Ranjan, C.~Castillo, and R.~Chellappa, ``L2-constrained softmax loss for
  discriminative face verification,'' \emph{arXiv preprint arXiv:1703.09507},
  2017.

\bibitem{Le2018}
N.~Le and J.~Odobez, ``Robust and discriminative speaker embedding via
  intra-class distance variance regularization,'' in \emph{Proceedings of
  Interspeech}, 2018, pp. 2257--2261.

\bibitem{Lazebnik2006}
S.~Lazebnik, C.~Schmid, and J.~Ponce, ``Beyond bags of features: Spatial
  pyramid matching for recognizing natural scene categories,'' in
  \emph{Proceedings of Computer Vision and Pattern Recognition (CVPR)}, 2006,
  pp. 2169--2178.

\bibitem{Sivic2003}
J.~Sivic and A.~Zisserman, ``Video google: A text retrieval approach to object
  matching in videos,'' in \emph{Proceedings of International Conference on
  Computer Vision (ICCV)}, 2003, pp. 1470--1477.

\bibitem{Kamishima2013}
Y.~Kamishima, N.~Inoue, and K.~Shinoda, ``Event detection in consumer videos
  using gmm supervectors and svms,'' \emph{EURASIP Journal on Image and Video
  Processing}, vol. 2013, pp. 1--13, 2013.

\bibitem{Hansen2015}
J.~Hansen and T.~Hasan, ``Speaker recognition by machines and humans: A
  tutorial review,'' \emph{IEEE Signal processing magazine}, vol.~32, no.~6,
  pp. 74--99, 2015.

\bibitem{Paszke2017}
A.~Paszke, S.~Gross, S.~Chintala, G.~Chanan, E.~Yang, Z.~DeVito, Z.~Lin,
  A.~Desmaison, L.~Antiga, and A.~Lerer, ``Automatic differentiation in
  pytorch,'' in \emph{Advances in Neural Information Processing Systems (NIPS)
  Autodiff Workshop}, 2017.

\bibitem{Shon2018}
S.~Shon, H.~Tang, and J.~Glass, ``Frame-level speaker embeddings for
  text-independent speaker recognition and analysis of end-to-end model,'' in
  \emph{Proceedings of Spoken Language Technology Workshop (SLT)}, 2018.

\bibitem{Yadav2018}
S.~Yadav and A.~Rai, ``Learning discriminative features for speaker
  identification and verification,'' in \emph{Proceedings of Interspeech},
  2018, pp. 2237--2241.

\bibitem{Chung2018}
J.~S. Chung, A.~Nagrani, and A.~Zisserman, ``Voxceleb2: Deep speaker
  recognition,'' in \emph{Proceedings of Interspeech}, 2018, pp. 1086--1090.

\end{thebibliography}


\end{document}